\documentclass[10pt,conference,letterpaper]{RWWTemplate}

\usepackage{times,amsmath,epsfig}
\usepackage{cite}

\newcommand\blfootnote[1]{%
		\begingroup
		\renewcommand\thefootnote{}\footnote{#1}%
		\addtocounter{footnote}{-1}%
		\endgroup
	}
	
\usepackage{subcaption}	
{\renewcommand{\baselinestretch}{0.94}

\title{Effect of Passive Reflectors for Enhancing Coverage of 28 GHz mmWave Systems in an Outdoor Setting}

\author{%
Wahab Ali Gulzar Khawaja{\small $~^{1}$}, Ozgur Ozdemir{\small $~^{1}$}, Fatih Erden{\small $~^{1}$}, Ismail Guvenc{\small $~^{1}$}, Martins Ezuma{\small $~^{1}$}\\ and Yuichi Kakishima{\small $~^{2}$}\vspace{6pt}\\
$~^{1}$Department of Electrical and Computer Engineering, North Carolina State University, Raleigh, NC\\
$~^{2}$DOCOMO Innovations, Inc., Palo Alto, CA \\
{\normalsize Email: \{wkhawaj, oozdemi, ferden, iguvenc, mcezuma\}@ncsu.edu, kakishima@docomoinnovations.com}
 }

\pdfoutput=1

\begin{document}
\maketitle

%
\begin{abstract}
The availability of large unused spectrum at millimeter wave~(mmWave) frequency bands has steered the future 5G research towards these bands. However, mmWave signals are attenuated severely in the non-line-of-sight~(NLOS) scenarios, thereby leaving the strong link quality by a large margin to line-of-sight~(LOS) links. In this paper, a passive metallic reflector is used to enhance the coverage for  mmWave signals in an outdoor, NLOS propagation scenarios. The received power from different azimuth and elevation angles are measured at 28~GHz in a parking lot setting. Our results show that using a 33~inch by 33~inch metallic reflector, the received power can be enhanced by 19~dB compared to no reflector case.    
\end{abstract}

\begin{keywords}
Coverage, mmWave, non-line-of-sight~(NLOS), outdoor, reflector.
\end{keywords}

\section{Introduction}
\blfootnote{This work has been supported in part by NASA under the Federal Award ID number NNX17AJ94A and by DOCOMO Innovations, Inc. We also thank Fujio Watanabe and Masato Takada from DOCOMO Innovations, Inc. for their valuable feedback in this study.}

There has been a surge in the use of smart communication devices in recent years. For example, in the U.S. the percentage of the population owning a smart phone has increased from $35\%$ in 2011 to $77\%$ in 2018~\cite{sp}. These smart devices can support high data rate applications that are becoming an essential part of the everyday life. However, due to increasing congestion at the sub-$6$~GHz spectrum, it is difficult to support high data rate applications in the future. This motivated the cellular industry to explore millimeter wave~(mmWave) frequency bands for mobile communications. Major hurdles for mmWave based system implementation include high free space attenuation, limited signal penetration through building structures, and small diffraction from large structural edges. This makes the radio signal planning for non-line-of-sight~(NLOS) paths very difficult.   

\begin{figure}[!t]
	\begin{subfigure}{0.33\textwidth}
	\centering
	\includegraphics[width=\textwidth, height = 4.4cm]{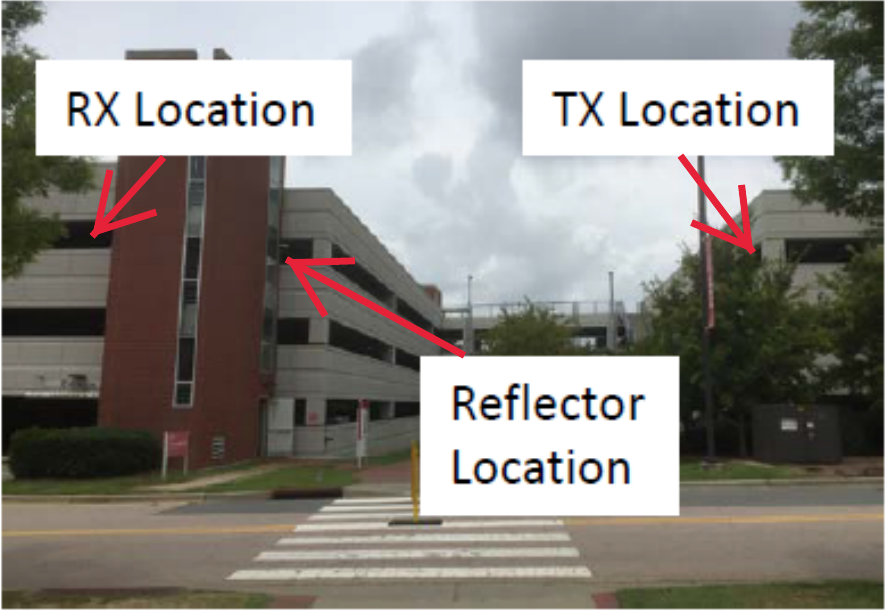}
	\caption{}
    \end{subfigure}			
	\begin{subfigure}{0.1\textwidth}
	\centering
    \includegraphics[width=\textwidth]{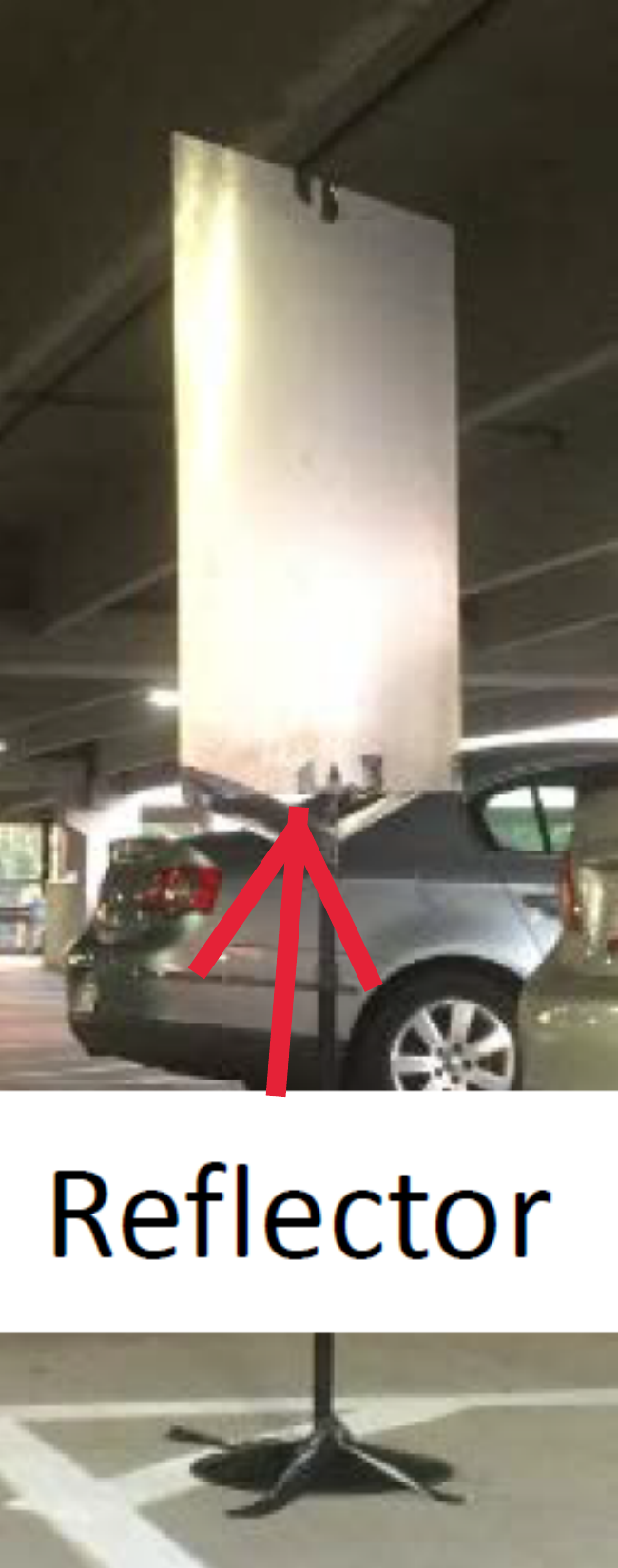}
	 \caption{}
     \end{subfigure}
     \caption{Measurement setup (a) Overview of the measurement setup (b) $33$ in $\times$ $33$ in metallic reflector.}\label{Fig:setup1}
     \vspace{-5mm}
\end{figure}

There are numerous solutions proposed in the literature in order to overcome the hurdle of high attenuation at the mmWave frequency bands, especially, for NLOS paths. These solutions include beam-forming and beam-steering techniques using multiple antennas, high transmit power and high sensitivity receivers, and use of multiple active repeaters. However, all of these solutions have limitations. Complex, expensive and high power consumption devices are required for beam-forming and beam-steering and it can still suffer in the NLOS propagation. Similarly, due to limitations on the transmit power emission by regulatory bodies, the transmit power can not be increased beyond a given value. Furthermore, using high sensitivity receivers and multiple active access points may not be economically and practically convenient.

\begin{figure*}[h!]
\centering
\centerline{\includegraphics[width=0.9\textwidth]{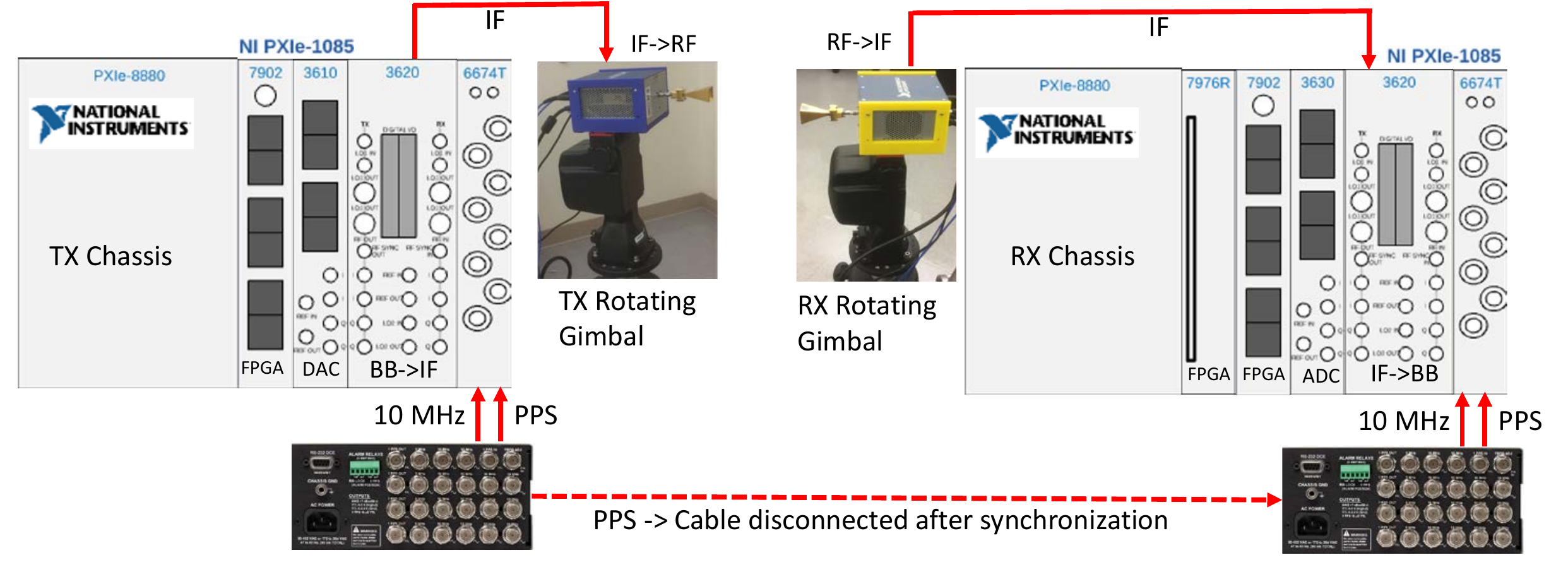}}
\caption{28 GHz channel sounder TX and RX hardware setup.}\label{Fig:pxisetup}\vspace{-3mm}
\end{figure*}

A feasible and more economical solution not extensively studied in the literature for mmWave signal enhancement in the NLOS scenarios is by using simple metallic passive reflectors in the propagation path. The properties of electromagnetic wave propagation are similar to the light~\cite{Light_EM}. Therefore, similar to light reflection principle, metallic reflectors can be used to reflect the electromagnetic waves. These electromagnetic wave reflections from metallic objects are better at higher frequencies due to smaller skin depth~\cite{reflection}, owing to lower material penetration. These metallic reflectors in the propagation link behave somewhat similar to a communication repeater. They also have the advantage of using no electricity and small initial investment, compared e.g. to wireless repeaters. Furthermore, they require negligible maintenance, have long life spans, and can also be part of common real life objects such as advertisement boards, lamp posts, and street signs.

There are limited studies available in the literature on using passive reflectors for downlink communications~\cite{Microwave_refl,Literature6}. Passive reflectors were mostly used in the past for point-to-point long distance links~\cite{NASA_refl,Literature4,Literature5} and is main focus of radar research studies~\cite{radar1,radar2, radar3}. In our recent work~\cite{wahab_indoor,hiranandani_vtc}, metallic passive reflectors of different shapes/sizes were used to enhance the coverage of NLOS scenarios in mmWave systems. A median gain of $20$~dB was observed with flat metallic reflector as compared to no reflector, whereas a cylindrical metallic reflector was found to provide more uniform coverage over the receiver grid. In~\cite{Literature3}, numerical analysis was carried out to observe the effect of a parabolic reflector placed on top of a building. The base station was placed on another building opposite to the reflector. The parabolic reflector was found to overcome the shadowed regions between the buildings. 

In this work we study the effect of reflectors on enhancing the coverage of mmWave systems in an outdoor setting as shown in Fig.~\ref{Fig:setup1}. A rotating gimbal is used to measure the received power from different directions in a parking lot. The measurement results show that  around $19$~dB power gain is achieved with the reflector, when compared to the communication link with no reflector.

\section{Hardware Setup}

The measurements were performed using NI mmWave PXI platforms at $28$~GHz~\cite{NImmwave}. The hardware setup is shown in Fig.~\ref{Fig:pxisetup}. 
The 10 MHz and pulse per second (PPS) signals generated by FS725 Rubidium~(Rb) clocks~\cite{SRS} are connected to PXIe 6674T modules at the transmitter and the receiver. The PPS output from one of the clocks is connected to the PPS input of the other clock so that the two clocks are synchronized before the carrying out the measurements. Once the clocks are synchronized, this connection can be removed so that the transmitter and the receiver can be separated from each other without any cable connecting them. 

Zadoff-Chu (ZC) sequence of length 2048 is periodically transmitted to sound the channel. The sounder has two modes of operation: 1~GHz and 2~GHz. In 2 GHz mode of operation which is used during these measurements, ZC sequence is over-sampled by 2 and filtered by root raised cosine (RRC) filter and the generated samples are uploaded into FPGA denoted by PXIe-7902. These samples are sent to PXIe-3610 digital to analog converter (DAC) with a sampling rate of $f_s=3.072$~GS/s. The PXIe-3620 module up-converts the base-band signal to IF and 28 GHz mmWave radio head up-converts the IF signal to RF. 

Directional horn antennas~\cite{sageM} are connected to the mmWave radio heads at the transmitter and receiver sides with 17 dBi gains and $26^{\circ}$ and $24^{\circ}$ beam-widths in the elevation and azimuth planes, respectively. The transmitter and receiver mmWave radio heads are placed on FLIR PTU-D48E gimbals~\cite{FlirSystems} in order to measure the angular profile of the channel.

At the receiver side, 28 GHz mmWave radio head down-converts the RF signal to IF. The IF signal is down-converted to base-band at the  PXIe-3620 module. The PXIe-3630 analog to digital converter module samples the base-band analog signal with the sampling rate of $f_s=3.072$~GS/s. The correlation and averaging operations are performed in PXIe-7976R FPGA operation and the complex channel impulse response (CIR) samples are sent to the PXIe-8880 host PC for further processing and saving to local disk. 

When the oversampling by two is performed, the channel sounder provides $2/f_s=0.65$~ns delay resolution in the delay domain. The analog to digital converter has around $60$~dB dynamic range and path loss up to $185$~dB can be measured. 

\subsection{Calibration for Non-Ideal Hardware Response}

\begin{figure}[!t]
	\begin{subfigure}{0.5\textwidth}
	\centering
	\includegraphics[width=0.9\textwidth]{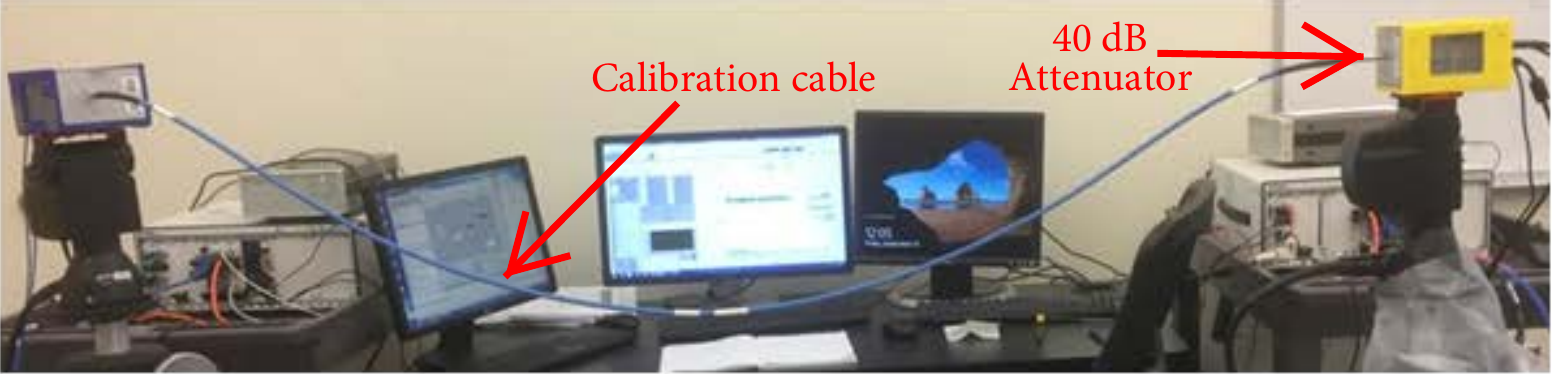} 
	\caption{}
    \end{subfigure}			
	\begin{subfigure}{0.23\textwidth}
	\centering
    \includegraphics[width=\textwidth]{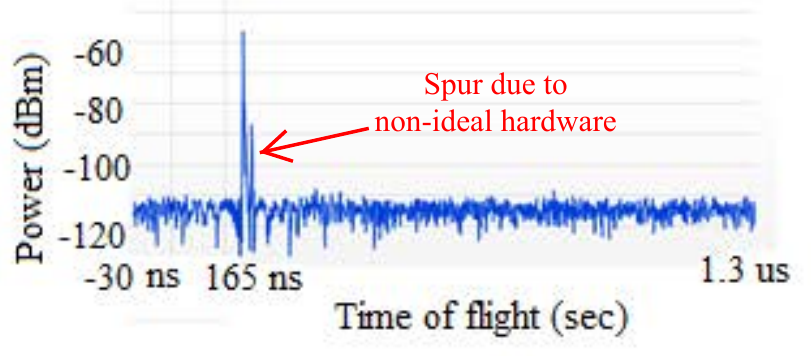}
	 \caption{}
     \end{subfigure}
     \begin{subfigure}{0.23\textwidth}
	\centering
    \includegraphics[width=\textwidth]{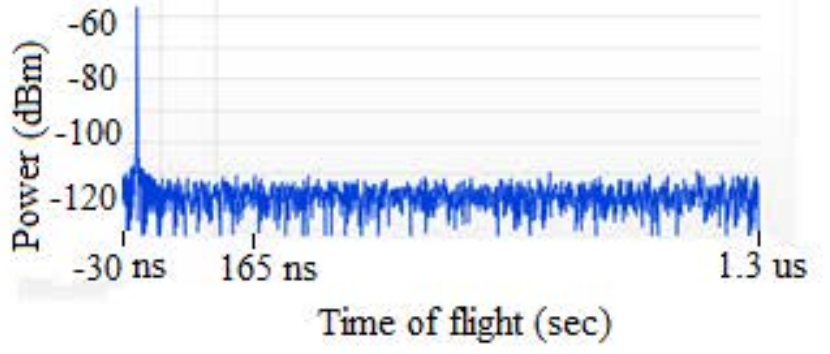}
	 \caption{}
     \end{subfigure}
     \caption{Calibration for hardware non-ideal hardware response, (a) Calibration cable connected between trasmitter and receiver mmWave radio heads (b) PDP obtained due to non-ideal hardware (c) PDP after calibration of the non-ideal hardware response}\label{Fig:calibration}
\end{figure}

One of the challenges when performing wide-band channel sounding is that the measurement hardware itself may introduce channel distortions which should be calibrated. One expects a flat response when a calibration cable is connected between the transmitter and the receiver as shown in Fig.~\ref{Fig:calibration}(a). However, due to the non-idealities of the hardware, spurs can be observed at the power delay profile (PDP) as seen in the example in Fig.~\ref{Fig:calibration}(b). By connecting the transmitter and the receiver with a calibration cable, it is possible to measure  and calibrate the non-flat frequency response of the hardware. During this measurement a 40~dB attenuator is used to protect the receiver when a signal with high power is transmitted. Fig.~\ref{Fig:calibration}(c) shows the response after calibration is performed where no spur is observed.

\section{Measurement Setup}

\begin{figure}[t!]
     \begin{subfigure}{0.2\textwidth}
	\centering
     \includegraphics[width=\textwidth]{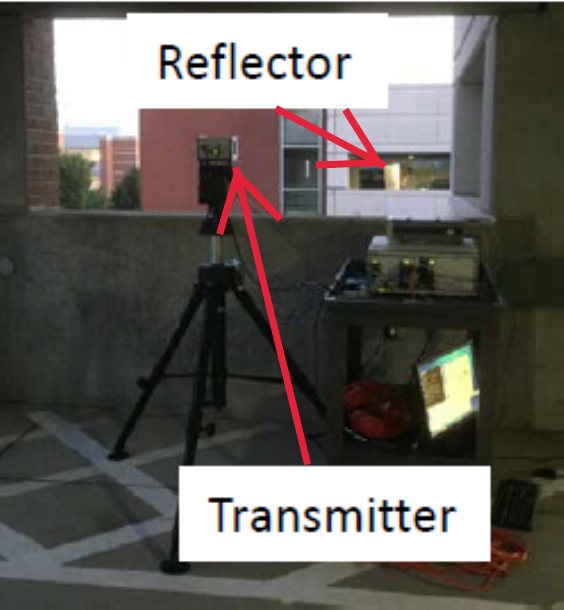}
	 \caption{}
     \end{subfigure}
      \begin{subfigure}{0.25\textwidth}
	\centering
    \includegraphics[width=\textwidth, height = 3.68cm]{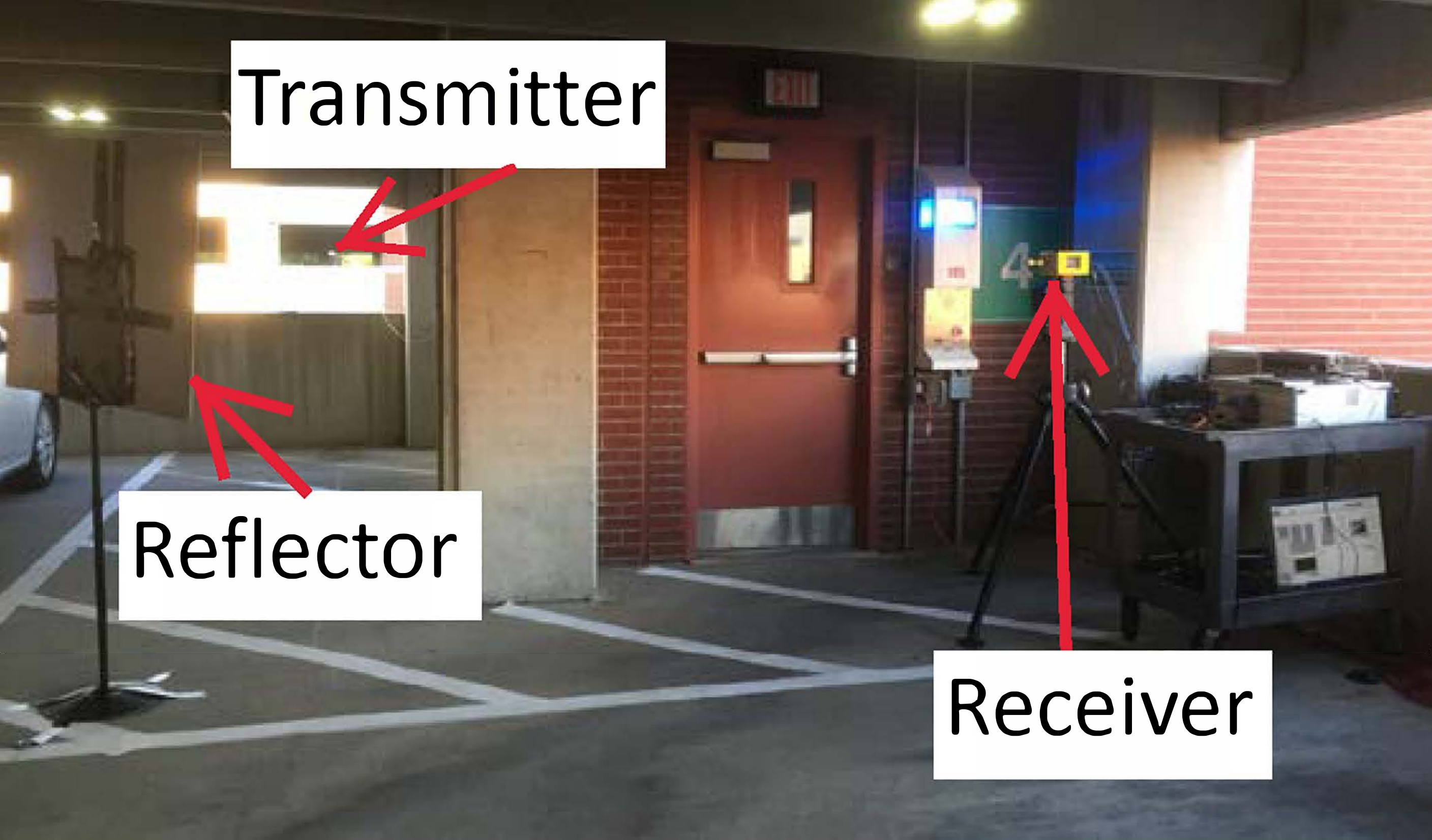}
	 \caption{}
     \end{subfigure}
    \caption{Measurement setup: (a) The transmitter shown on the gimbal; during the measurements, the transmitter was not rotated. (b) Receiver location; the reflector is positioned at $45^{\circ}$ angle to maximize the received power.}\label{Fig:setup2}
    \vspace{-3mm}
\end{figure}

\begin{figure}[t!]
\centering
\centerline{\includegraphics[width=0.5\textwidth]{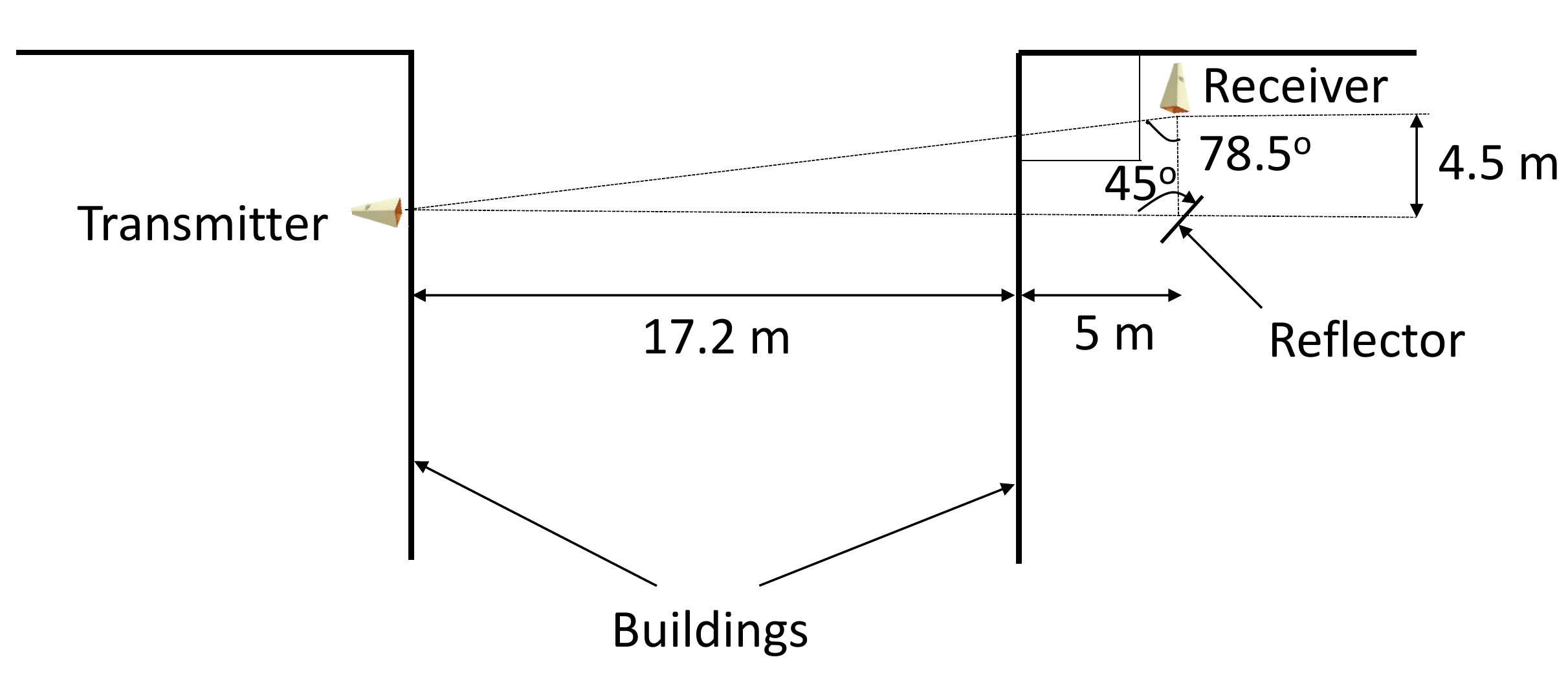}}
\caption{Top view of the layout of the measurement scenario with the heights of the transmitter and receiver at $1.5$~m and the reflector center is aligned to the center of the transmitter/receiver antenna.}\label{Fig:layout}
\end{figure}

The measurements were performed inside two parking buildings next to each other at North Carolina State University (NCSU) campus as shown in~Fig.~\ref{Fig:setup2}. The receiver is located behind brick walls surrounding the stairs so that there is no direct line of sight (LOS) path between the transmitter and the receiver. 
A passive metallic reflector of size $33$~inch by $33$~inch is used 4.5~m away from the receiver as shown in Fig.~\ref{Fig:layout}. The receiver antenna is mounted on a rotatable gimbal in order to collect energy from different directions.

\begin{figure}[!t]
\begin{center}
	\begin{subfigure}{0.4\textwidth}
	\centering
	\includegraphics[width=\textwidth]{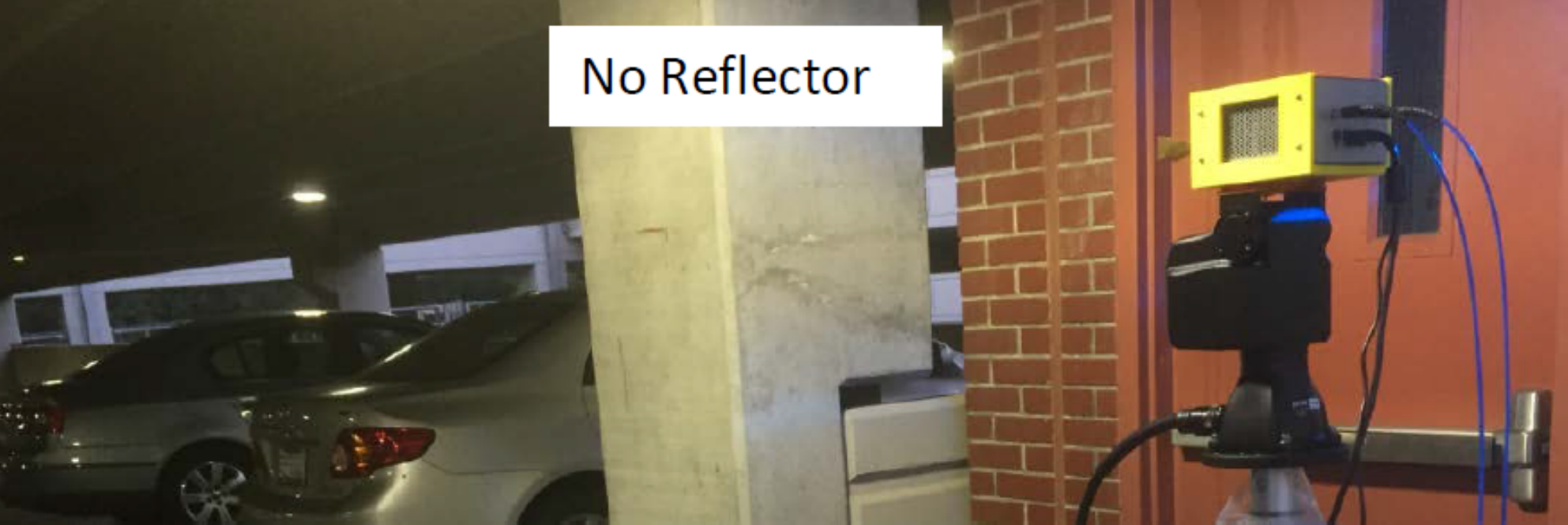}
	\caption{}
    \end{subfigure}			
	\begin{subfigure}{0.4\textwidth}
	\centering
    \includegraphics[width=\textwidth]{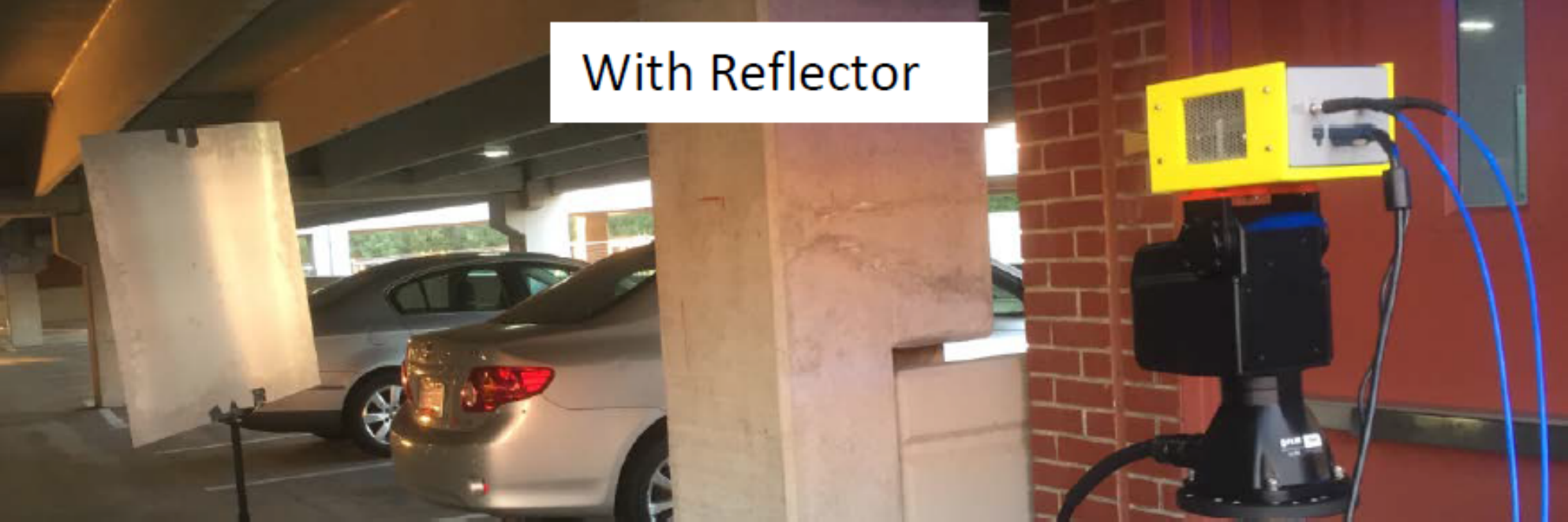}
	 \caption{}
     \end{subfigure}
\end{center}\vspace{-2mm}
    \caption{Scenarios with: (a) no reflector, (b) with reflector.}\label{Fig:scenario}\vspace{-2mm}
\end{figure}

\begin{figure}[t!]
\centering
\centerline{\includegraphics[width=0.5\textwidth]{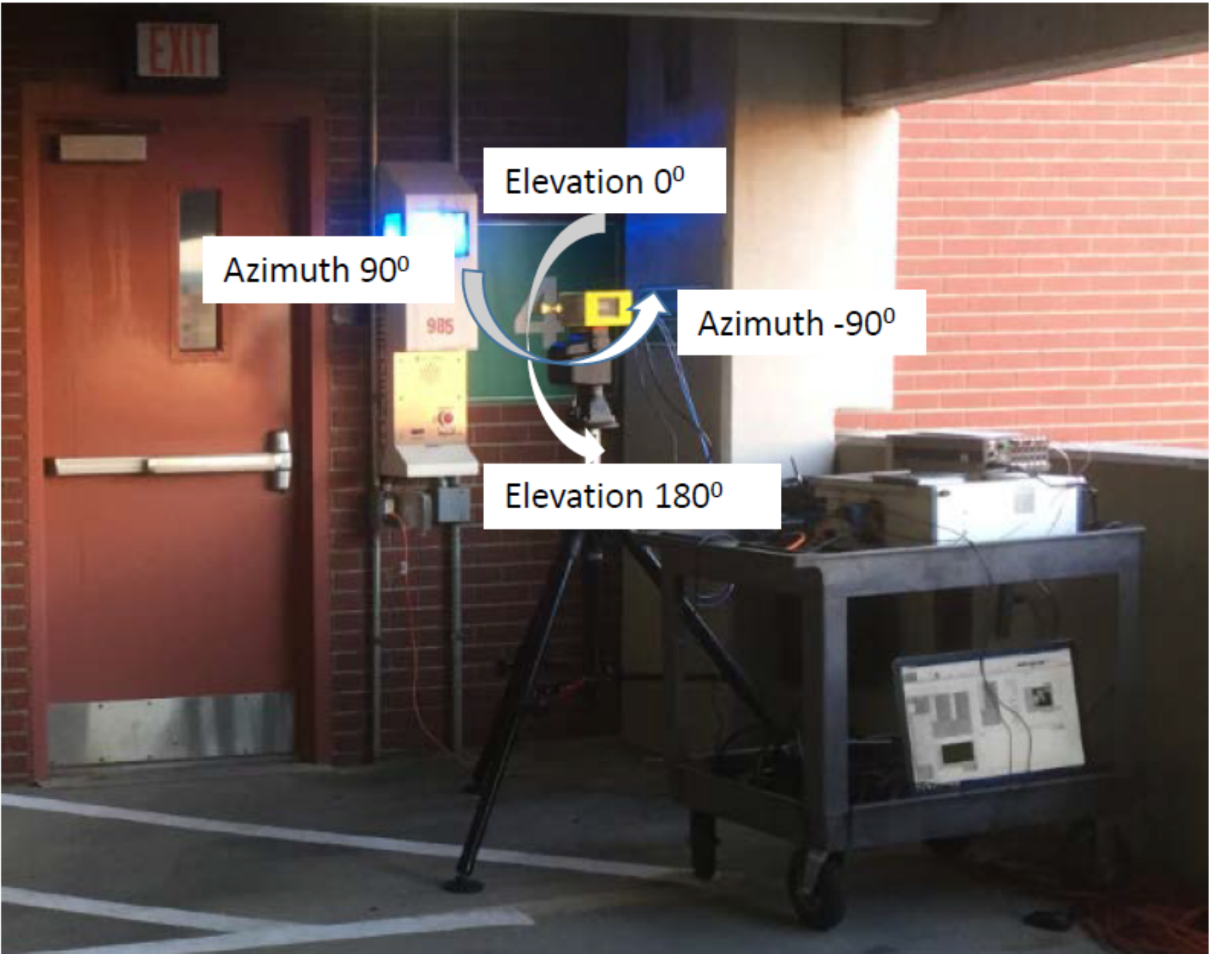}}
\caption{The orientation of elevation and azimuth angles}\label{orientation:fig}
\vspace{-3mm}
\end{figure}

For comparison we consider two scenarios: one without reflector and one with reflector as shown in Fig.~\ref{Fig:scenario}.  The transmitter is not rotated and its position is fixed at $90^{\circ}$ elevation angle \emph{facing the location of the reflector}. We adjust the heights of the transmitter and receiver equal to each other so that it is expected that the received power will be maximum when the receiver is at $90^{\circ}$ elevation angle as shown in Fig.~\ref{orientation:fig}. The gimbal at the receiver side scans the azimuth plane from $-165^{\circ}$ to $165^{\circ}$ with $10^{\circ}$ increments and the elevation plane from $1^\circ$ to $119^\circ$ with $10.7^\circ$ increments.  

\section{Measurement Results}
\begin{figure}[t!]
\begin{center}
\begin{subfigure}{0.5\textwidth}
\centering
\includegraphics[width=\textwidth]{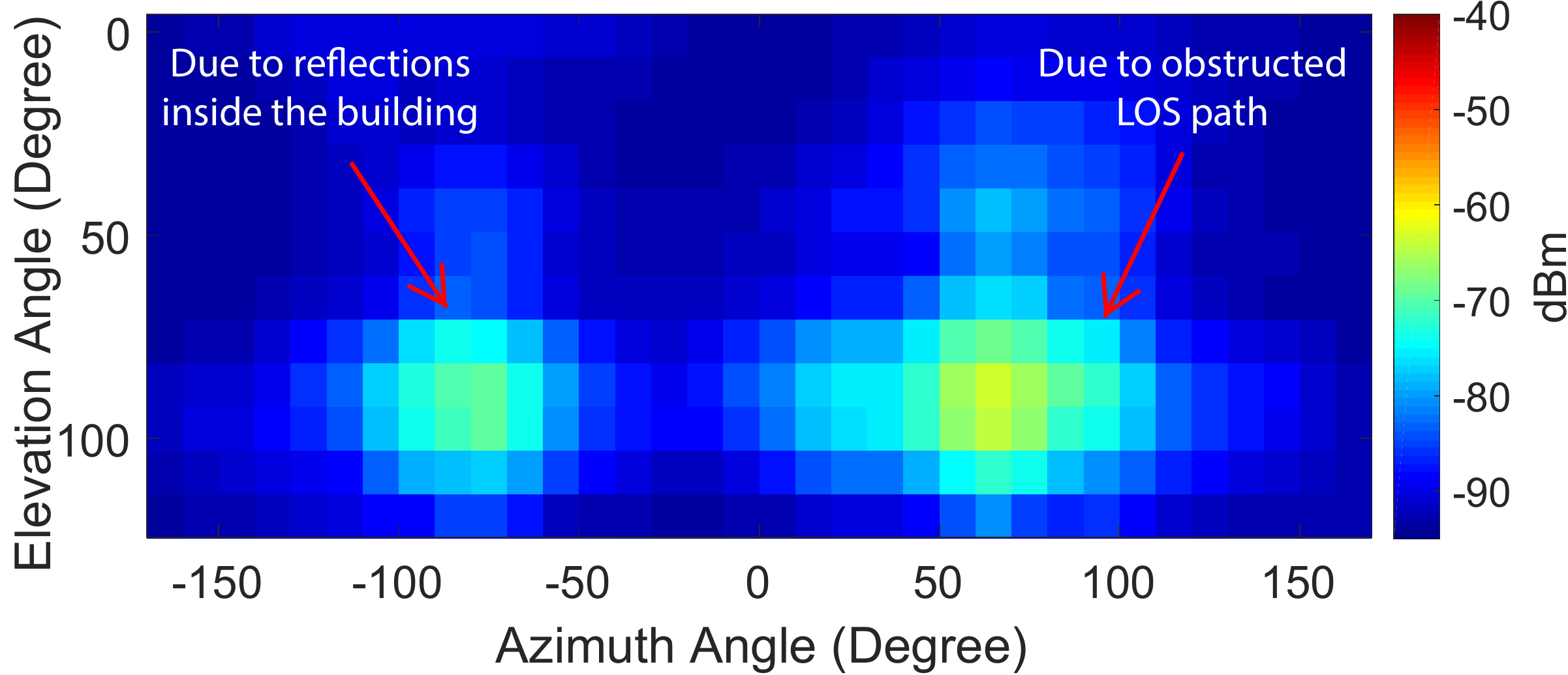}
\caption{}
\end{subfigure}	

\begin{subfigure}{0.5\textwidth}
\centering
\includegraphics[width=\textwidth]{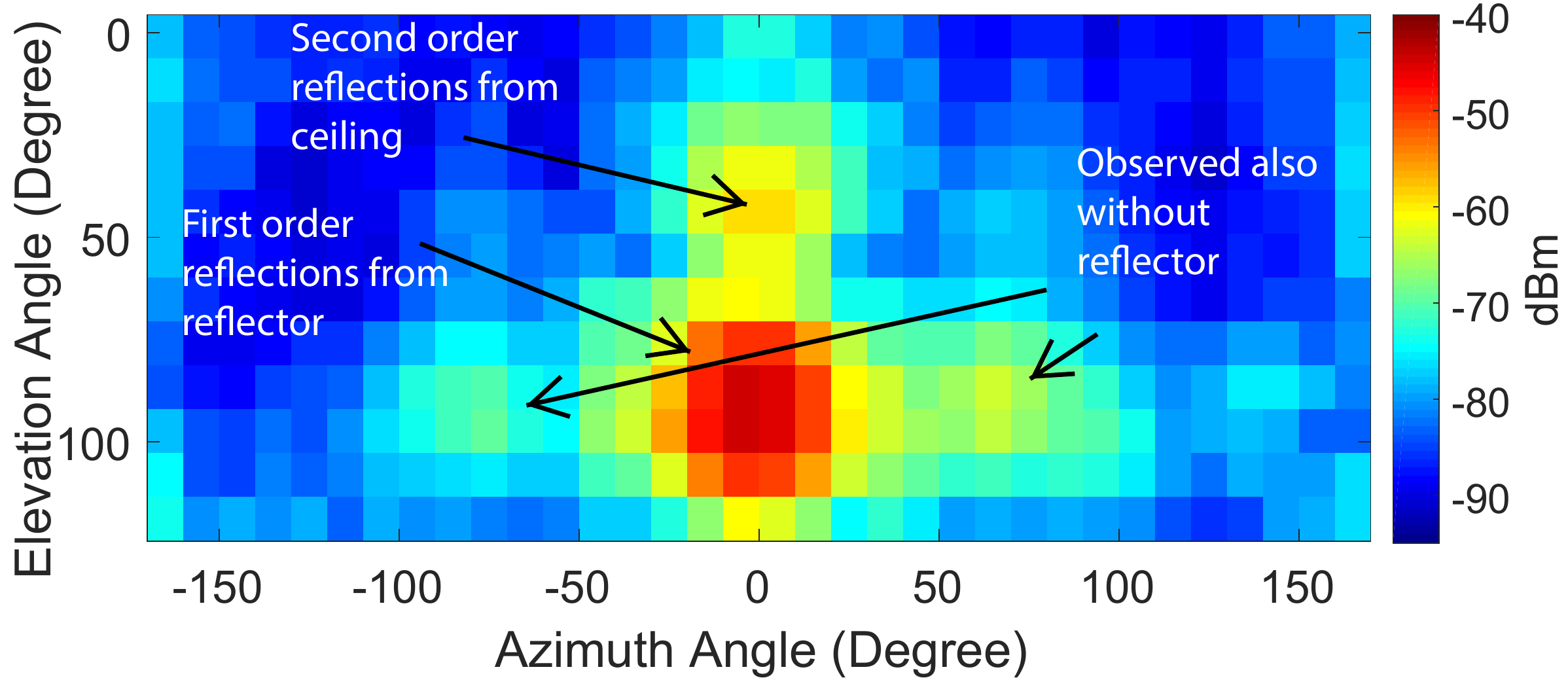}
\caption{}
\end{subfigure}

\begin{subfigure}{0.5\textwidth}
\centering
\includegraphics[width=\textwidth]{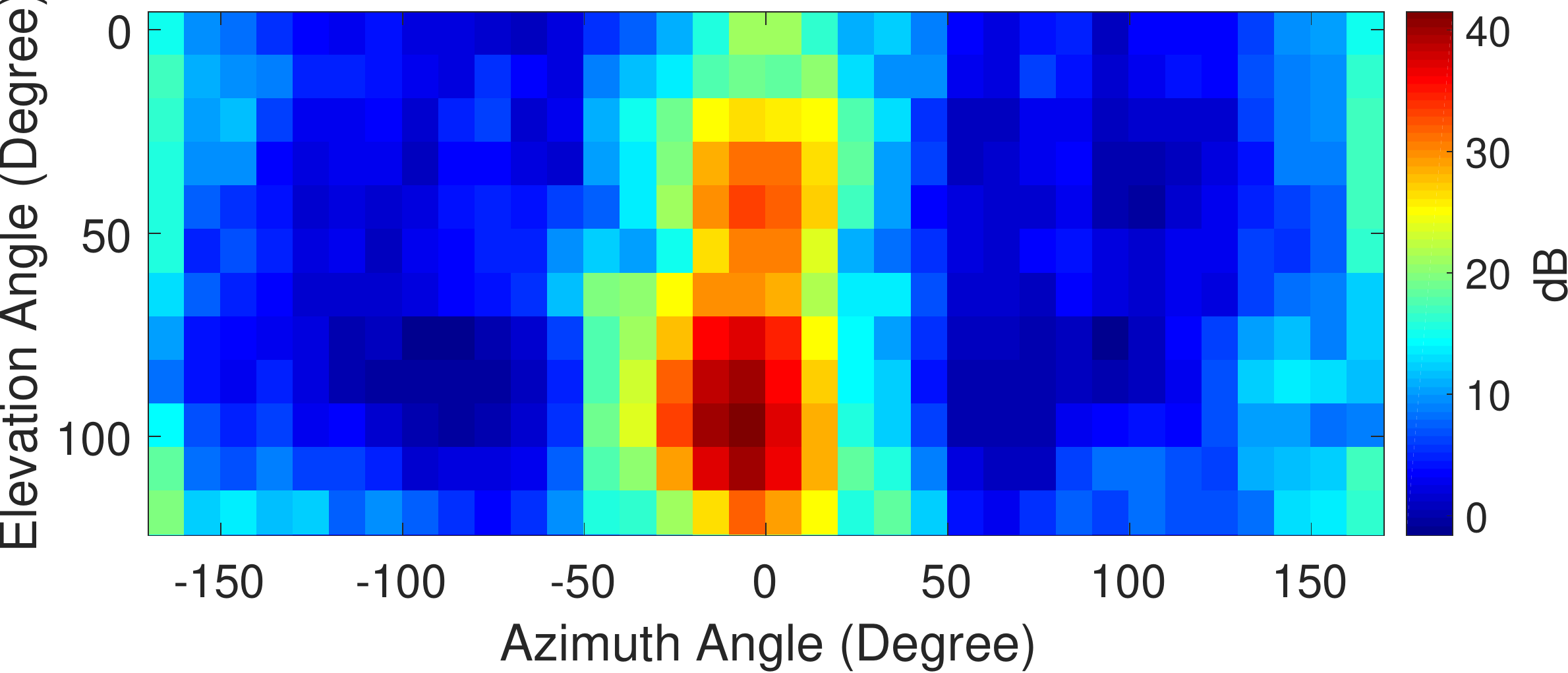}
\caption{}
\end{subfigure}
\end{center}
\caption{Measurement results for (a) without reflector, (b) with reflector, (c) the gain obtained with reflector compared to no reflector case.} \label{Fig:Results}
\end{figure}

The transmit power is set to $20$~dBm during measurements. For each antenna position the total received power is measured by summing the power values from the PDP provided by the channel sounder. 
Fig.~\ref{Fig:Results}(a), shows the measured power when there is no reflector. In this case the maximum power which is -64 dBm is observed at the azimuth angle of $65^\circ$ and an elevation angle of $87^{\circ}$. Furthermore, the signal is received with -70 dBm power at the same elevation angle and at $-75^\circ$ azimuth angle. The strongest path corresponds to the direct path between the transmitter and the receiver and the other path corresponds to some reflections inside the building.

Fig.~\ref{Fig:Results}(b) shows the measured power with the reflector in place. The two strong components when there is no reflector shows up in this case as well. However, here we observe the strongest received power of -45 dBm at azimuth angle of  $-5^\circ$ and the same elevation angle of $87^\circ$. This is the first order reflection from the reflector. A second order reflection from the ceiling having a power of -59 dBm, is also observed at the elevation angle $44^\circ$ and the same azimuth angle $-5^\circ$. We observe that when the reflector is used, the strongest power is increased by (-45 dBm) - (-64 dBm) = 19 dB.

Finally, Fig.~\ref{Fig:Results}(c) plots the power gain obtained with reflector compared to without reflector by subtracting the results in Fig.~\ref{Fig:Results}(a) from the results in Fig.~\ref{Fig:Results}(b). We note that at the position where the receiver faces the reflector, around 40 dB gain is observed.

\section{Conclusions}
In this work, we have studied the effect of a passive metallic reflector to enhance the coverage for mmWave signals in an outdoor setting. The measurement results show that using $33$~in$\times$$33$~in metallic reflector, $19$~dB gain in power is possible compared to no reflector case. Our future work includes extensive indoor/outdoor measurement campaigns to characterize the channel scattering properties for typical indoor/outdoor 5G communication environments.   





\begin{thebibliography}{10}
\providecommand{\url}[1]{#1}
\csname url@samestyle\endcsname
\providecommand{\newblock}{\relax}
\providecommand{\bibinfo}[2]{#2}
\providecommand{\BIBentrySTDinterwordspacing}{\spaceskip=0pt\relax}
\providecommand{\BIBentryALTinterwordstretchfactor}{4}
\providecommand{\BIBentryALTinterwordspacing}{\spaceskip=\fontdimen2\font plus
\BIBentryALTinterwordstretchfactor\fontdimen3\font minus
  \fontdimen4\font\relax}
\providecommand{\BIBforeignlanguage}[2]{{%
\expandafter\ifx\csname l@#1\endcsname\relax
\typeout{** WARNING: IEEEtran.bst: No hyphenation pattern has been}%
\typeout{** loaded for the language `#1'. Using the pattern for}%
\typeout{** the default language instead.}%
\else
\language=\csname l@#1\endcsname
\fi
#2}}
\providecommand{\BIBdecl}{\relax}
\BIBdecl

\bibitem{sp}
\BIBentryALTinterwordspacing
``Mobile fact sheet, pew research center,'' accessed: 7-30-2018. [Online].
  Available: \url{http://www.pewinternet.org/fact-sheet/mobile/}
\BIBentrySTDinterwordspacing

\bibitem{Light_EM}
M.~Kerker, \emph{The scattering of light and other electromagnetic
  radiation}.\hskip 1em plus 0.5em minus 0.4em\relax Elsevier, 2016.

\bibitem{reflection}
S.~N. Ghosh, \emph{Electromagnetic theory and wave propagation}.\hskip 1em plus
  0.5em minus 0.4em\relax CRC Press, 2002.

\bibitem{Microwave_refl}
Y.~Huang, N.~Yi, and X.~Zhu, ``Investigation of using passive repeaters for
  indoor radio coverage improvement,'' in \emph{IEEE Ant. Propag. Society
  Sympos.}, vol.~2, June 2004, pp. 1623--1626 Vol.2.

\bibitem{Literature6}
J.~L. D. L.~T. Barreiro and F.~L.~E. Azpiroz, ``Passive reflector for a mobile
  communication device,'' Aug. 2006, {US} Patent 7,084,819.

\bibitem{NASA_refl}
C.~C. Cutler, ``Passive repeaters for satellite communication systems,'' Feb.~9
  1965, uS Patent 3,169,245.

\bibitem{Literature4}
J.~L. Ryerson, ``Passive satellite communication,'' \emph{Proc. of the IRE},
  vol.~48, no.~4, pp. 613--619, April 1960.

\bibitem{Literature5}
Y.~E. Stahler, ``Corner reflectors as elements passive communication
  satellites,'' \emph{IEEE Trans. Aerospace}, vol.~1, no.~2, pp. 161--172, Aug.
  1963.

\bibitem{radar1}
J.~Bjornholt, G.~Hamman, and S.~Miller, ``Electronic fence using
  high-resolution millimeter-wave radar in conjunction with multiple passive
  reflectors,'' ~15 2002, uS Patent 6,466,157.

\bibitem{radar2}
W.~Khawaja, K.~Sasaoka, and I.~Guvenc, ``{UWB} radar for indoor detection and
  ranging of moving objects: An experimental study,'' in \emph{Proc. IEEE Int.
  Workshop Ant. Technol. (iWAT)}, 2016, pp. 102--105.

\bibitem{radar3}
C.~Bredin, J.-M. Goutoule, R.~Sanchez, J.-P. Aguttes, and T.~Amiot, ``High
  resolution {SAR} micro-satellite based on passive reflectors,'' in
  \emph{Proc. IEEE Int. Geoscience and Remote Sensing Symposium}, vol.~2, 2004,
  pp. 1196--1199.

\bibitem{wahab_indoor}
W.~Khawaja, O.~Ozdemir, Y.~Yapici, I.~Guvenc, and Y.~Kakishima, ``Coverage
  enhancement for {mmWave} communications using passive reflectors,'' in
  \emph{Proc. IEEE Global Symp. Millimeter Waves ({GSMM})}, Boulder, Colorado,
  May 2018.

\bibitem{hiranandani_vtc}
S.~Hiranandani, S.~Mohadikar, W.~Khawaja, O.~Ozdemir, I.~Guvenc, and
  D.~Matolak, ``Effect of passive reflectors on the coverage of {IEEE} 802.11ad
  {mmWave} systems,'' in \emph{Proc. IEEE Vehic. Technol. Conf. (VTC)
  workshops}, Chicago, IL, Aug. 2018.

\bibitem{Literature3}
Z.~Peng, L.~Li, M.~Wang, Z.~Zhang, Q.~Liu, Y.~Liu, and R.~Liu, ``An effective
  coverage scheme with passive-reflectors for urban millimeter-wave
  communication,'' \emph{IEEE Ant. Wireless Propag. Lett.}, vol.~15, pp.
  398--401, 2016.

\bibitem{NImmwave}
\BIBentryALTinterwordspacing
{National Instruments}, ``{mmWave} {T}ransceiver {S}ystem,'' accessed:
  7-31-2018. [Online]. Available: \url{http://www.ni.com/sdr/mmwave/}
\BIBentrySTDinterwordspacing

\bibitem{SRS}
\BIBentryALTinterwordspacing
{Standford Research Systems}, ``{FS725} {R}ubidium {F}requency {S}tandard,''
  accessed: 7-31-2018. [Online]. Available:
  \url{https://www.thinksrs.com/products/fs725.html}
\BIBentrySTDinterwordspacing

\bibitem{sageM}
\BIBentryALTinterwordspacing
{SAGE Millimeter, Inc}, ``{SAR-1725-34KF-E2} {R}ectangular {H}orn {A}ntenna,''
  accessed: 7-31-2018. [Online]. Available:
  \url{https://www.sagemillimeter.com/17-dbi-gain-wr-34-2-92-mm-female-connector-rectangular-horn-antenna/}
\BIBentrySTDinterwordspacing

\bibitem{FlirSystems}
\BIBentryALTinterwordspacing
{Flir Systems}, ``{FLIR PTU-D48E} {P}an/{T}ilt {U}nit,'' accessed: 7-31-2018.
  [Online]. Available: \url{https://www.flir.com/products/ptu-d48e/}
\BIBentrySTDinterwordspacing

\end{thebibliography}


\end{document}